\documentclass[]{interact}
\usepackage[utf8]{inputenc}
\usepackage{amsmath,amsfonts,amssymb}
\usepackage{float}
\usepackage{graphicx}
\usepackage[caption=false]{subfig}
\usepackage{physics}
\usepackage{hyperref}

\usepackage[numbers,sort&compress]{natbib}
\bibpunct[, ]{[}{]}{,}{n}{,}{,}
\renewcommand\bibfont{\fontsize{10}{12}\selectfont}
\makeatletter
\def\NAT@def@citea{\def\@citea{\NAT@separator}}
\makeatother

\hypersetup{
    bookmarksopen=true,
    colorlinks=true,
    linkcolor=blue,
    citecolor=blue,
    urlcolor=blue,
    pdftitle={Directionality on periodic materials},
    pdfauthor={N. Guarin-Zapata, C. Valencia, J. Gomez},
    pdfkeywords={Periodic materials, Dispersion relation, Bloch analysis},
    pdfsubject={Directionality on periodic materials},
    pdfpagemode=UseOutlines,
    pdfstartview=FitH
}

\graphicspath{{./img/}}

\title{Analysis of the directionality on periodic materials
\footnote{This is an original manuscript of an article published by Taylor \&
Francis in Mechanics of Advanced Materials and Structures on 02 Jul 2023,
available at: \url{https://doi.org/10.1080/15376494.2023.2226958}}}

\author{\name{Nicolás Guarín-Zapata\textsuperscript{a}
    \thanks{CONTACT: Nicolás Guarín-Zapata. Email: \url{nguarinz@eafit.edu.co}},
    Camilo Valencia\textsuperscript{a}, Juan Gomez\textsuperscript{b}
    \footnote{Juan Gómez was formerly at Universidad EAFIT}}
    \affil{\textsuperscript{a}Universidad EAFIT, School of Applied Sciences and
    Engineering, Medellín 050022, Colombia\\
    \textsuperscript{b}Risk and Design Consulting, Medellín 050022, Colombia}}

\begin{document}

\maketitle

\setcounter{footnote}{0}

\begin{abstract}
There is an increasing interest in the study of metamaterials and periodic
materials across disciplines. These are anisotropic and their properties present
directionality. For example, the wave speed depends on the propagation direction.
Furthermore, they are heterogeneous, and their directionality depends on their
spectra. Common approaches to describe anisotropy have been used in the
large-wavelength approximation corresponding to static properties. Here we
present an anisotropy measure based on the dynamic behavior. It receives
dispersion surfaces from Bloch analyses and outputs a curve/surface with bulk
directionality encoded on it. We present results for elastodynamics, but it is
applicable to other phenomena.
\end{abstract}


\section{Introduction}

Across disciplines, there is a growing interest in the research and design of
metamaterials and periodic materials \cite{book:metamaterials-banerjee,
book:photonics-molding_flow, book:deymier2013}. The emergence of unusual
properties such as effective negative mass, negative refraction, negative Poisson
ratio, and acoustic/electromagnetic cloaking has piqued the attention of
scientists \cite{norris2012, hussein2014, goldsberry2018}. As a result, there is
a trend in designing microstructures to control bulk properties \cite{milton1995,
norris2011metal, willis2016, valencia2019, guarin-zapata_bandgap_2019}. Although
direct numerical simulations of the bulk of the material and the entire
microstructure are feasible \cite{sigalas2000}, the most common method is to
take advantage of the material's periodicity and model a single cell \cite{pennec2010, 
hussein2014}. This is accomplished using Bloch's theorem \cite{bloch1929}.
Bloch's theorem was developed in solid-state physics \cite{book:brillouin2003,
book:kittel}, but it has since been applied to electrodynamics
\cite{book:photonics-molding_flow}, acoustics \cite{zhang2004, 
book:deymier2013}, and elastodynamics \cite{hussein2014, 
valencia_general-purpose_2019, guarin-zapata_finite_2020}.

Periodic materials are anisotropic because of discrete symmetries in their unit
cells \cite{book:nye1985, book:kittel, moakher2006closest, 
maurin2018probability}. This anisotropy behavior translates into directionality 
in wave propagation, i.e., the speed of waves depends on the direction
of propagation. If we want to study the propagation of waves
in a periodic material, we are interested in characterizing this directionality.
Furthermore, due to the heterogeneous nature of periodic
materials, their anisotropy depends on the frequency of the waves propagating
through them. That is, we could talk of a \emph{dynamically induced anisotropy}. This
behavior has been reported in the literature \cite{ruzzene2003wave, ruzzene2005directional}
and is related with the interaction of the waves and the microstructure. Mediated
by the relation between the wavelengths and characteristic length in the
material. We illustrate this phenomenon with a 2D mass-spring lattice, where
we can compute the solution analytically.

There have been some efforts in defining descriptors that quantify 
the anisotropy level of materials \cite{zener1948elasticity, kanit2006, 
ranganathan2008}, they are not perfect because they reduce all the information 
to a single number and do not provide information about the preferred direction
of propagation or symmetries present. Guarín-Zapata et al. 
\cite{guarin-zapata_bandgap_2019} used a qualitative approach to 
compare anisotropy for different trasnsversely isotropic materials and select 
them to tune the behavior of helicoidal composites. This was
used for transversely isotropic materials, where the dispersion relations can be
obtained analytically \cite{auld1973} and used to analyze a single
layer of a composite. For heterogeneous materials and materials with other
symmetries besides isotropy and transverse isotropy, this type of information
is not available analytically and numerical simulations are required.
Thus, a general tool to analyze the bulk anisotropy of materials is desired.

Here we present a new anisotropy measure
based on the dynamic behavior of the material. Our approach takes as input
dispersion (hyper-) surfaces obtained using a Bloch analysis and outputs a
curve/surface with the bulk directionality encoded on it. We present some
results for the case of elastodynamics, but they are directly applicable to
problems with other underlying physical phenomena. As such, this method allows
extending directionality analyses in metamaterials and periodic materials beyond
the long wavelength regime.

This paper introduces a novel anisotropy measure based on the dynamic behavior
of the material. As such, it considers the material's heterogeneity and
works beyond the long-wavelength regime. The method can be applied to a specific
frequency or any chosen frequency range. The starting point is dispersion
(hyper-) surfaces, which are a common result when applying Bloch's theorem
to periodic materials. We begin with some generalities about periodic
materials and Bloch's theorem, which is used to describe bulk behavior with a
single unit cell. Then, we describe the propagation of waves in anisotropic
elastic materials in two and three dimensions. Following this section, we
present the suggested method for analyzing periodic material directionality,
as well as some tests for the method in analytically and numerically obtained
dispersion relations. Although all the examples presented are from the area of
elastodynamics, this method applies to any material that employs dispersion
(hyper-) surfaces as input, such as electrodynamics or quantum mechanics.

\section{Periodic materials}

A periodic material is defined by the spatial repetition of a given motif in 
one, two, or three dimensions. The motifs refer to heterogeneities in the 
material properties at the microstructural level and can contain different
materials, topologies, and shapes. For example, in 
the case of electromagnetic waves, we have \emph{photonic crystals} and the 
periodicity of electric permittivity and magnetic permeability 
\cite{book:photonics-molding_flow}. For elastic waves, the term is 
\emph{phononic crystals} and we have periodicity in the stiffness and mass 
density of the material \cite{book:deymier2013}. 
Such periodic materials are described by a lattice
and by an elementary unit cell. Figures 1(a)–1(c) show a three-dimensional
material with periodicity in one, two, and three dimensions. A set of base
vectors defines the lattice (Fig. 1(d)). These allow the construction of the
whole material through successive applications of translation operations of the
unit cell. To study this type of material it is
common to take advantage of the periodicity of the material properties and
express the solution using Bloch's theorem
\cite{book:brillouin2003}.
\begin{figure}
    \centering
    \subfloat[]{\includegraphics[width=0.22\textwidth]{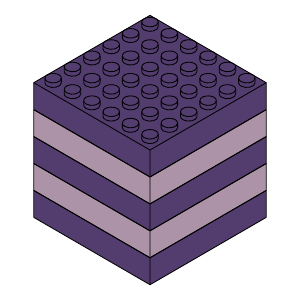}}
    \subfloat[]{\includegraphics[width=0.22\textwidth]{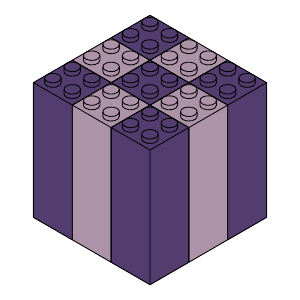}}
    \subfloat[]{\includegraphics[width=0.22\textwidth]{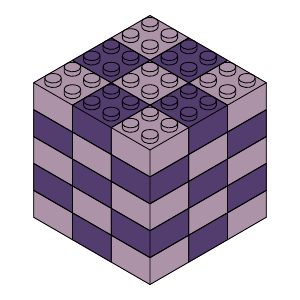}}
    \subfloat[]{\includegraphics[width=0.32\textwidth]{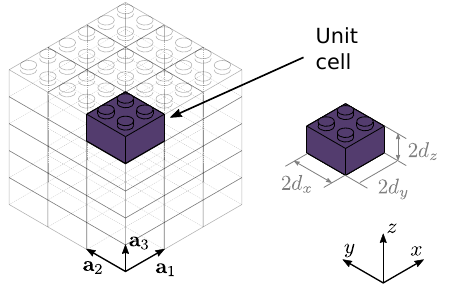}}
    \caption{3D material with different periodicities. Regardless of the space 
    dimensionality its periodicity could be in (a) one (b) two or (c) three 
    space dimensions. (d) The purple lego-brick defines the fundamental unit cell 
    which allows construction or filling of space after applying translation 
    operations according to the lattice vector \(\mathbf{a}\).}
    \label{fig:lego}
\end{figure}

\subsection{Bloch's theorem}
Let us consider a generalized wave equation of the form
\begin{equation}
\mathcal{L} \vb{u}(\vb{x}) = -\omega^2\vb{u}(\vb{x})\, ,
\label{eq:gen_wave}
\end{equation}
where \(\mathcal{L}\) is a positive definite operator \cite{algebraic_waves, 
book:reddy_functional_analysis}, \(\vb{u}\) is the field of interest, and 
\(\omega\) is the angular frequency. Bloch's theorem establishes that
solutions to \eqref{eq:gen_wave} are of the form
\begin{equation}
\vb{u}(\vb{x}) = \vb{w}(\vb{x}) e^{i\boldsymbol{\kappa}\cdot\vb{x}}\, ,
\label{eq:bloch}
\end{equation}
where \(\mathbf{w}(\mathbf{x})\) is a function with the same 
periodicity of the material. Opposite sides of the unit cell
are separated by a vector \(\mathbf{a}\) and, as a consequence,
\[\vb{u}(\vb{x} + \vb{a}) = \vb{u}(\vb{x})e^{i\boldsymbol{\kappa}\cdot\vb{a}}\, .\]
The expressions \(\mathbf{u}(\mathbf{x} + \mathbf{a})\) and \(\mathbf{u}(\mathbf{x})\)
give the field at \(\mathbf{x} + \mathbf{a}\) and \(\mathbf{x}\), while  
\(\mathbf{a} = \mathbf{a}_1 n_1 + \mathbf{a}_2 n_2 + \mathbf{a}_3 n_3\) is the 
lattice translation vector shown in Figure \ref{fig:lego}(d). The term 
\(e^{i\boldsymbol{\kappa}\cdot\mathbf{a}}\) represents a phase shift between opposite 
sides of the unit cell. This relationship between opposite sides of the 
fundamental cell stated in the theorem through the boundary terms permits the 
characterization of the fundamental properties of the material with the 
analysis of a single cell.

\begin{figure}[h]
    \centering
    \includegraphics[width=4 in]{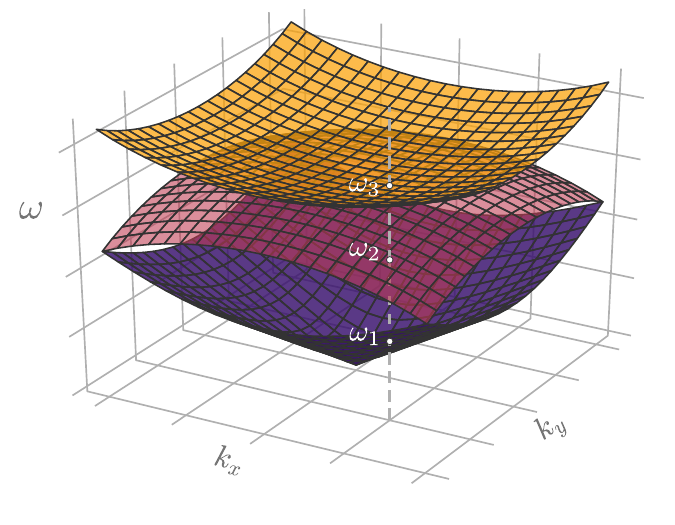}
    \caption{Schematic depicting the multiple branches obtained in a Bloch
    analysis. For a single point in the wavevector space there are multiple
    values for the frequency.}
    \label{fig:schematic_branches}
\end{figure}

Computationally, equation \eqref{eq:gen_wave} is commonly translated to a
generalized eigenvalue problem through a numerical method such as the Finite
Element Method. And we end up with a system of the form
\[[K]\{\vb{U}\} = \omega^2[M] \{\vb{U}\}\, ,\]
where \([K]\) is the stiffness matrix and \([M]\) is the mass matrix. Bloch's
theorem can be applied through boundary conditions imposed strongly by 
directly including the phase shifts at the element level or performing row and
column operations in these matrices \cite{valencia_general-purpose_2019}; or
weakly imposing them through Lagrange multipliers or penalty methods
\cite{michel1999effective, sukumar_bloch-2009}. After this process, we end
up with the following system
\[[K_R(\boldsymbol{\kappa})]\{\vb{U}_R\} = \omega^2[M_R(\boldsymbol{\kappa})] \{\vb{U}_R\}\, ,\]
where  \(\mathbf{k}\) is the wave vector which, is progressively assigned
successive values, in such a way that the first Brillouin zone is fully covered.
Each evaluation for a particular wave vector and the solution of the related
eigenvalue problem yields tuples of the form \((\mathbf{k},\omega_n)\)
representing a plane wave propagating at frequency \(\omega_n\). The subindex
\(n\) refers to the \(n\)th eigenvalue computed for the input wave vector
\(\boldsymbol{\kappa}\); this is depicted in Figure \ref{fig:schematic_branches}.

Having multiple dispersion surfaces is common for problems in electrodynamics or
elastodynamics \cite{book:photonics-molding_flow, book:metamaterials-banerjee}.
This is a consequence of the vector nature of the physical quantities of
interest: electric/magnetic field or displacement vector, respectively.
In the case of homogeneous materials, we have two transverse modes for
electromagnetic waves and two transverse modes plus a longitudinal one for
elastic waves. Nevertheless, Bloch analysis introduces an additional
complication since the computations are done in the first Brillouin zone. Here,
wavevectors are determined up to an additive constant that is written in terms
of the lattice vectors
\[\vb{a}_1 n_1 + \vb{a}_2 n_2 + \vb{a}_3 n_3\quad \forall n_1, n_2, n_3
\in \mathbb{Z}\, .\]

\begin{figure}[h]
    \centering
    \includegraphics[width=6 in]{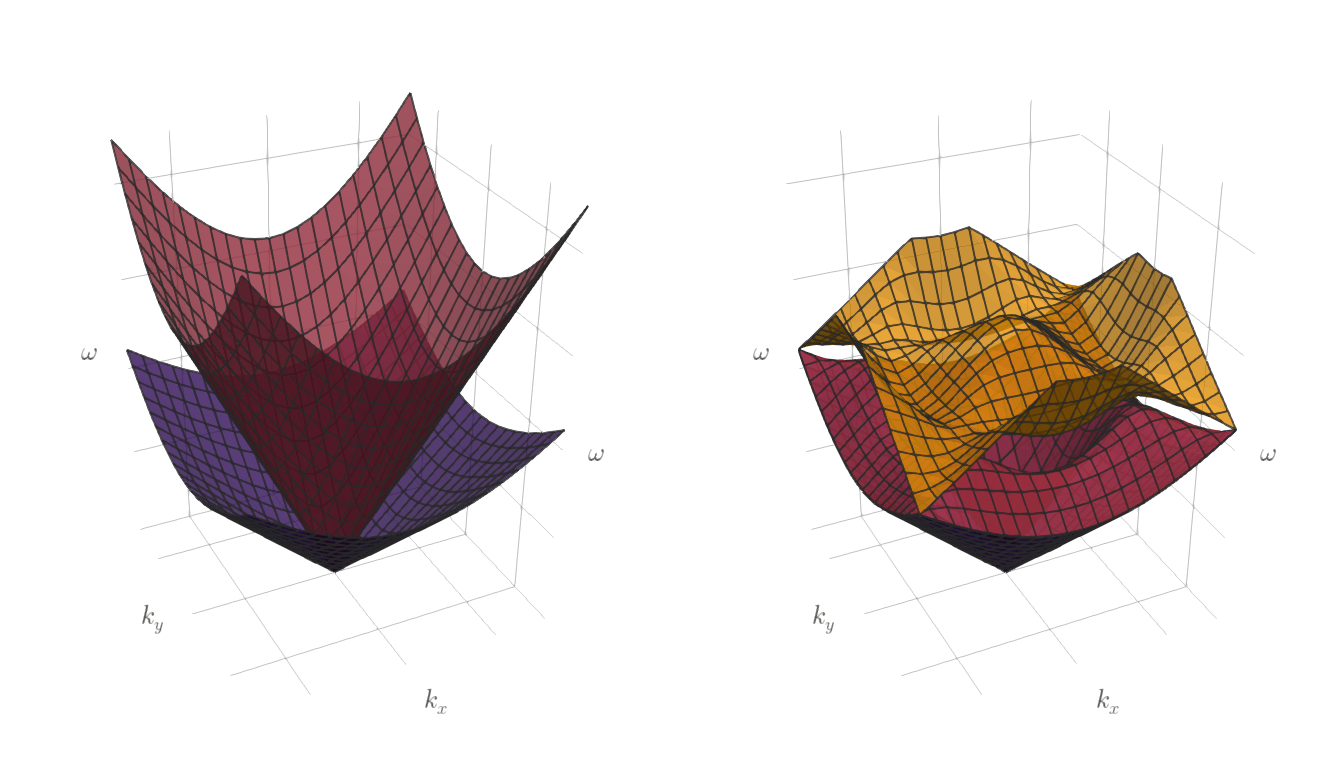}
    \caption{Comparison of the dispersion surfaces for a homogeneous material
    computed with the classical expression (left) with the dispersion surfaces
    obtained using Bloch analysis (right). Here we are showing the first
    three branches.}
    \label{fig:disp_folded}
\end{figure}

This means that we map the vectors \(\boldsymbol{\kappa}\) and
\(\boldsymbol{\kappa} + \vb{a}_1 n_1 + \vb{a}_2 n_2 + \vb{a}_3 n_3\)  to the same point in
the first Brillouin zone. Figure \ref{fig:disp_folded} illustrates this effect
for the first three dispersion surfaces in the case of a homogeneous material.

\begin{figure}[h]
    \centering
    \includegraphics[width=4.5 in]{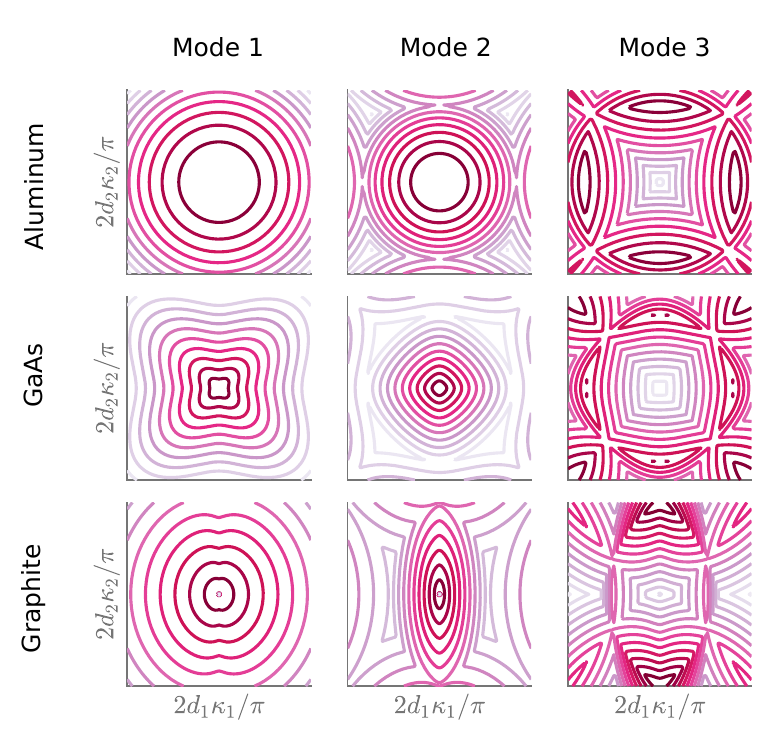}
    \caption{Comparison of the contours of iso-frequency for the first three
    dispersion modes for isotropic, cubic and orthotropic materials.
    \textbf{(Top)} Contours for aluminum, an isotropic example.
    \textbf{(Middle)} Contours for GaAs, a cubic example. \textbf{(Bottom)}
    Contours for graphite, an orthotropic example.}
    \label{fig:apparent_aniso}
\end{figure}

As shown in Figure \ref{fig:disp_folded}, when using Bloch analysis, we obtain
dispersion surfaces ordered by frequency. This might lead to conclude that
materials are anisotropic when they are not, as illustrated in Figure
\ref{fig:apparent_aniso}. For low wavenumbers, we can see that the material is
isotropic/anisotropic when looking at the first mode. Nevertheless, this is not
the case when looking at the second and third modes. Furthermore, anisotropy in
one material might be more pronounced in one propagation mode than in others.

\begin{figure}
    \centering
    \includegraphics[width=5 in]{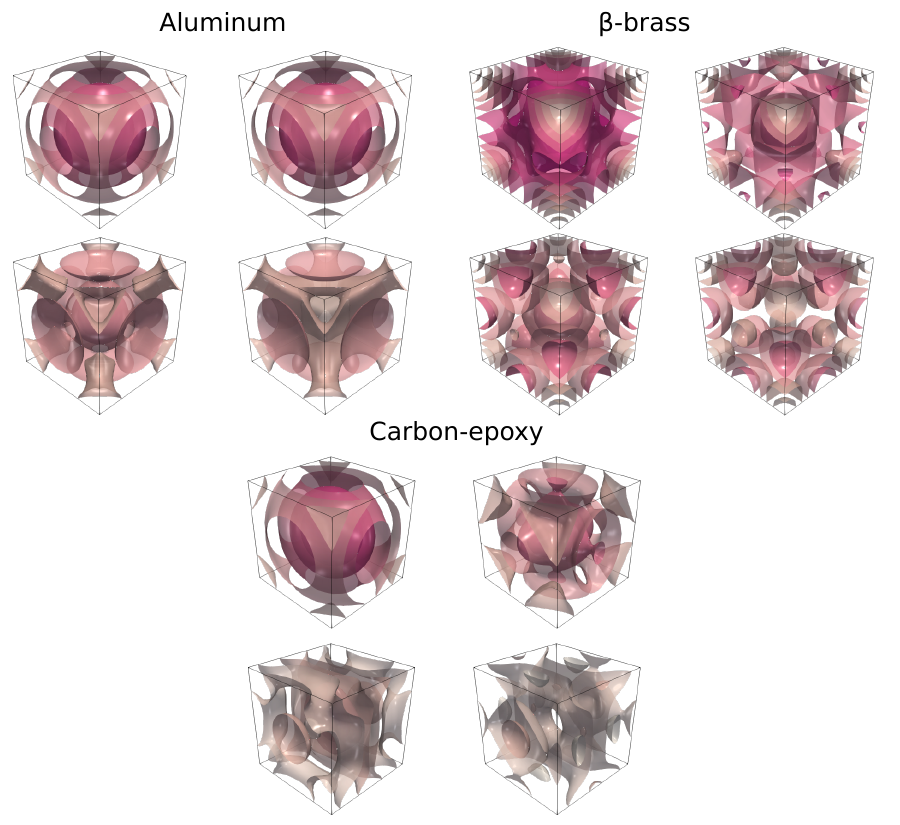}
    \caption{Comparison of the the first four dispersion modes for isotropic,
    cubic, and transverse isotropic materials. The surfaces represent
    iso-frequency values for the dispersion relations.}
    \label{fig:apparent_aniso3d}
\end{figure}

This apparent anisotropy is more difficult to analyze if we consider wave
propagation in three dimensions instead of two, as can be seen in Figure
\ref{fig:apparent_aniso3d}.

\section{Elastic wave propagation in anisotropic media}

As mentioned before, periodic materials are inherently anisotropic due to their
microstructure. Their \emph{level} of anisotropy obeys the symmetries present.
In this section we describe the propagation of waves for elastic
solids of general anisotropy.

Let us consider a wave that propagates in a solid. In this case, the
conservation of density of momentum in the absence of body forces is written as
\[\sigma_{ij,i} = \rho\pdv[2]{u_i}{t}\, ,\]
with \(\sigma_{ij}\) the stress tensor, \(u_i\) the displacement vector, and
\(\rho\) the mass density. In the case of a linear elastic material, Hooke's law
is given by
\[\sigma_{ij} = c_{ijkl} u_{k, l}\, ,\]
where \(c_{ijkl}\) is the stiffness tensor that can have up to 21 constants in
the case of general anisotropy.

If we assume a plane-wave solution of the form
\[u_j = U_j e^{i\kappa(n_m x_m - v_p t)}\, ;\]
where \(\kappa=|\boldsymbol{\kappa}|\) is the wavenumber, \(n_m\) is a unit
vector in the direction of \(\boldsymbol{\kappa}\), and \(v_p\) is the phase speed; we
end up with the Christoffel wave equation \cite{buchwald1959, auld1973,
book:carcione2007}:
\[[\Gamma_{ij} - \rho v_p^2\delta_{ij}]U_j = 0\, ,\]
where \(\Gamma_{ij} = c_{ijkl} n_k n_l\) is the Christoffel stiffness tensor or
acoustic tensor, and \(\delta_{ij}\) is the Kronecker delta. This is an
eigenvalue problem with eigenvalues \(\rho v_p^2\). Here, \(v_p\) represents
the phase speeds of the material. The corresponding characteristic polynomial is
\begin{equation}\label{eq:christo}
    \det[\Gamma_{ij} - \rho v_p^2\delta_{ij}] = 0\, .
\end{equation}

\begin{figure}[h]
    \centering
    \includegraphics[width=4 in]{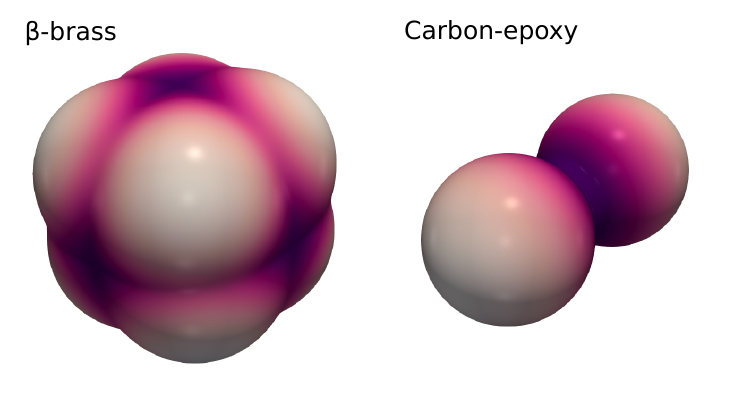}
    \caption{Comparison of the directionality of the phase speed for
    quasi-longitudinal (qP) modes for a cubic (left) and
    transverse isotropic material (right).}
    \label{fig:phase_speed}
\end{figure}

From \eqref{eq:christo} we conclude that there are three propagating waves,
each one with a different polarization. These directions are orthogonal since
\(\Gamma_{ij}\) is symmetric. Of the three polarizations, we have two 
quasi-transverse (qS) waves and one quasi-longitudinal (qP) wave. Where the
prefix \emph{quasi} implies that two of the modes are close to orthogonal to
the wavevector and the other is close to parallel to it. As we can see, the
value of \(\Gamma_{ij}\) depends on the direction of propagation implying that
the phase speed depends on the direction of the wave. This
effect is represented in Figure \ref{fig:phase_speed}, where we are visualizing
the phase speed for two anisotropic materials: \(\beta\)-brass and carbon-epoxy.

We can rewrite \eqref{eq:christo} as
\begin{equation}\label{eq:christo_freq}
    \Omega(\omega, \boldsymbol{\kappa}) = \det\left[\Gamma_{ij} - 
\rho \frac{\omega^2}{\Vert\boldsymbol{\kappa}\Vert^2}\delta_{ij}\right]\, ,
\end{equation}
considering that the phase speed is defined as
\(v_p^2 = \omega^2/\Vert\boldsymbol{\kappa}\Vert^2\). The group velocity is defined as
\cite{auld1973}
\begin{equation}\label{eq:group_vel}
\vb{v}_g = \frac{\nabla_{\boldsymbol{\kappa}} \Omega}{\partial \Omega/\partial \omega}\, ,
\end{equation}
and for lossless media it represents the direction of energy flow.

\section{Evaluation of directionality}

Elastic anisotropy plays a role in different applications such as
metallurgy \cite{zener1948elasticity}, geophysics \cite{thomsen1986},
wave propagation in composites \cite{guarin-zapata_bandgap_2019}, and
metamaterials \cite{casadei2013anisotropy}, among others. Thus, for a particular
application, one might need a material with more or less anisotropy, and the
question of how anisotropic different materials are arises naturally.

\subsection{Dynamically induced anisotropy}

As mentioned before, in periodic material their anisotropy depends on the
frequency of the waves propagating through them  \cite{ruzzene2003wave,
ruzzene2005directional}. That is, there exists a dynamically induced anisotropy
that relates to the interaction of the waves and the microstructure. We can
contrast this case with the static case. Which corresponds to the
large-wavelength approximation of the dynamic case.

\begin{figure}[h]
\centering    
\includegraphics[width=4 cm]{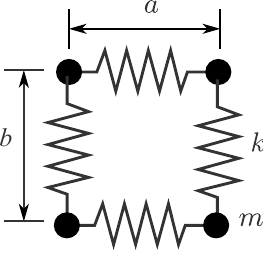}
\caption{Unit cell for a 2D mass spring lattice. Here \(k\) is the stiffness
of the spring, \(m\) is the mass, \(a\) is the distance between neighbors in
the \(x\)-direction, and \(b\) is the distance between neighbors in the
\(y\) direction.}
\label{fig:spring-mass}
\end{figure}

To illustrate this phenomenon, here we present as an example the dispersion
for a 2D mass-spring lattice. Figure \ref{fig:spring-mass} presents the unit
cell for this lattice.

\begin{figure}[h]
\centering    
\includegraphics[width=4 cm]{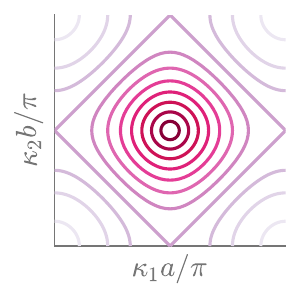}
\caption{Iso-frequency contours for the dispersion relation in the spring-mass
lattice.}
\label{fig:disp_spring-mass}
\end{figure}

The analytic solution, in this case, is the
following --- see \cite{hussein2014} for a deduction of this solution:
\[\Omega^2 = 2[2 - \cos(\kappa_1 a/\pi) - \cos(\kappa_2 b/\pi)]\, ,\]
where \(\Omega = \omega/\omega_0\) is the normalized frequency,
\(\omega_0 = \sqrt{k/m}\) is the frequency for a single spring-mass system,
\(\kappa_1\) is the wave number in the \(x\)-direction, \(\kappa_2\) is the
wave number in the \(y\)-direction, \(a\) is the distance between neighbors in
the \(x\)-direction, and \(b\) is the distance between neighbors in the
\(y\) direction. Figure \ref{fig:disp_spring-mass} presents the dispersion
relation for this problem.

\begin{figure}[h]
\centering    
\includegraphics[width=5 in]{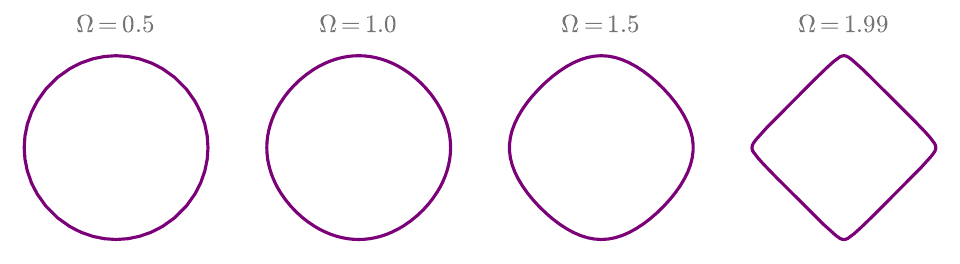}
\caption{Comparison of the directionality for a spring-mass lattice for
different normalized frequencies \(\Omega = 0.5\), \(\Omega = 1.0\),
\(\Omega = 1.5\), and \(\Omega = 1.99\).}
\label{fig:disp_contours_spring-mass}
\end{figure}

If we analyze the directionality for this problem, we see that for low frequency
it seems to be isotropic, but it starts to turns anisotropic when the 
frequency is changed. This is depicted in Figure \ref{fig:disp_contours_spring-mass}
We have different normalized frequencies: \(\Omega = 0.5\), \(\Omega = 1.0\),
\(\Omega = 1.5\), and \(\Omega = 1.99\). For this particular
problem, the maximum propagating frequency is \(\Omega = 2\).

\subsection{Some existing measures of anisotropy}

One of the first measures of anisotropy was the Zener ratio
\cite{zener1948elasticity} which is defined for cubic materials and is written
as
\[a_r = \frac{2 C_{44}}{C_{11} - C_{12}}\, .\]
From an intuitive perspective, it is defined as the ratio between the classical
shear modulus and the \emph{new one} that appears in cubic materials. The 
Zener ratio is one for isotropic materials. One extension of the Zener ratio
for materials with more general anisotropy replaces the elastic coefficients by
averages while retaining the form of the original ratio \cite{kanit2006}
\[a_\text{gen} = \frac{2Y_{44}}{Y_{11} - Y_{12}}\, ,\]
with
\[Y_{11} = \frac{C_{11} + C_{22} + C_{33}}{3}\,,
Y_{12} = \frac{C_{12} + C_{23} + C_{13}}{3}\, ,
Y_{44} = \frac{C_{44} + C_{55} + C_{66}}{3}\, .\]

There are other proposed metrics for measuring anisotropy
such as the ratio of the maximum and minimum phase speed for the
quasi-transverse modes \cite{ledbetter2006} or the norm of the projection
to the closest isotropic tensor \cite{book:carcione2007}.
Ranganathan and Ostoja-Starzewski \cite{ranganathan2008} compute some of these
metrics for several crystals and compare them with a new metric termed universal
anisotropy index (UAI). Although these metrics seem to be usable for general
anisotropic materials they were conceived with homogeneous materials in mind.

Regarding the anisotropy of composites, it is common to present the
directionality as group velocity polar histograms based
on iso-frequency contours \cite{ruzzene2005directional, ruzzene2003wave}.
Furthermore, Casadei and Rimoli \cite{casadei2013anisotropy} computed an
anisotropy index considering each propagation mode separately. This measure is,
essentially, the standard deviation for each wave propagation mode. The main
drawback with these approaches is that they considered just the first two or
three modes of propagation, that is, low frequencies. Since phononic crystals
can present dispersion, it is expected that the anisotropy (directionality)
depends on the frequency.

\subsection{Method for directionality}

\begin{figure}[h]
    \centering
    \includegraphics[width=5 in]{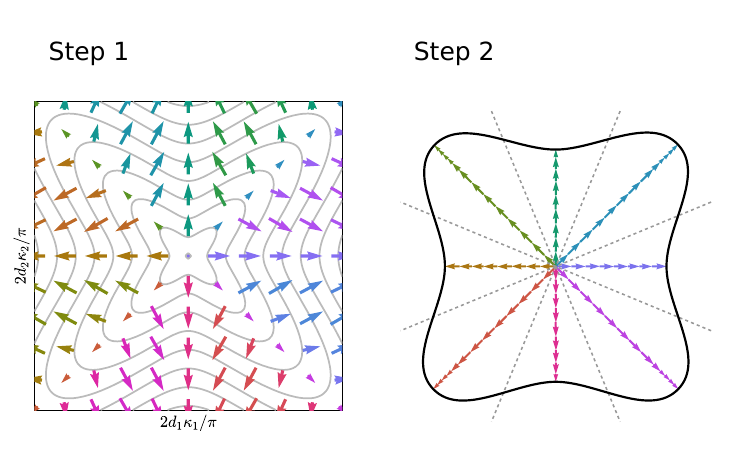}
    \caption{In \textbf{Step 1} we compute the gradient of the the \(i\)th
    mode of the dispersion curve. Then, in \textbf{Step 2} the group velocity
    vectors are rearranged according to their direction and added. The resulting
    envelope curve represents the directionality for that mode.}
    \label{fig:direct_method_schematic}
\end{figure}

Valencia et al. \cite{valencia2019} proposed a method to characterize the directionality
of phononic crystals that considers the contribution of multiple modes, not just
the low-frequency ones as in previous works \cite{ruzzene2005directional,
ruzzene2003wave}. Thus, the approach allows a more complete description of the
directional response in a material and is valid in the low and high-frequency
regimes. They defined the wave propagation directionality, \(D\), as
\begin{equation}\label{eq:dir}
D = \sum_{\substack{i\\e>tol}}d_i(\theta)\, , \quad d_i(\theta) =
    C(\nabla M_i)
\end{equation}
where \(M_i\) is the \(i\)th mode in the dispersion relation, \(\nabla M_i\) is
its gradient with respect to the wavevector \(\boldsymbol{\kappa}\), \(tol\) is a 
predefined tolerance, \(\theta = [0,2\pi]\) is the angle defining the
propagation direction. In this definition, the operator \(C\) associates each
vector to its direction (\(\theta\)) and adds it to the previous vector sharing 
the same direction. Consequently, \(d_i = C\left( \nabla M_i \right)\) 
corresponds to a weighted polar histogram representing the distribution of 
group velocity for mode \(M_i\) in any propagation direction. The weight is 
given by the number of times a particular direction appears in the (discrete) 
dispersion relations. We can summarize this method as:
\begin{itemize}
\item
Start from the frequency surface and compute the gradient to obtain group
velocities.

\item
Sort the (discrete) gradient by angle and add their magnitudes when they share
the angle.
\end{itemize}
These steps are depicted in Figure \ref{fig:direct_method_schematic}.

\begin{figure}[h]
    \centering
    \includegraphics[width=4 in]{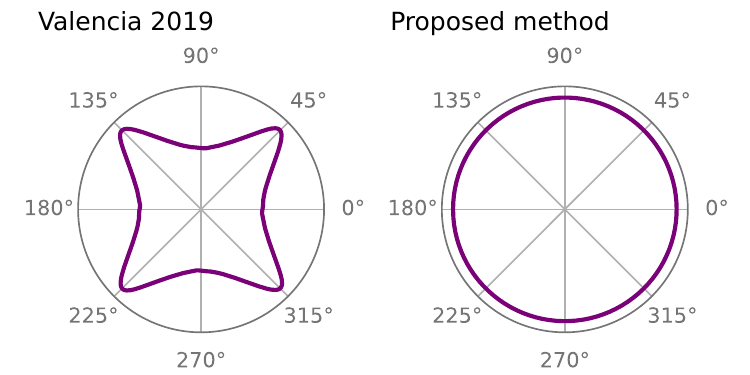}
    \caption{Comparison of the computed directionality for the first dispersion
    branch in a homogeneous material using (\textbf{left}) \cite{valencia2019}
    method and (\textbf{right}) our approach.}
    \label{fig:direct_method_comparison}
\end{figure}

One problem with this method is that it considers group velocity vectors for
regions with wavevectors of different magnitudes. For example, when
considering a square unit cell it will include more vectors associated to
directions \((1, 1)\), \((-1, 1)\), \((-1, -1)\) and \((1, -1)\).
This makes the method dependent on the choice of the unit cell shape.
Our proposed improvement is to average vectors instead of accumulating them.
This is similar to restricting the analysis to wavevectors enclosed
by a sphere with constant wavevector magnitude---in solid state physics
states enclosed by a constant-energy surface are considered to compute the
density of states \cite{book:kittel}. As an example, let us compare the
computed directionality for the first dispersion branch. That is, the first
eigenvalue obtained from the Bloch analysis of the unit cell. This is shown in Figure
\ref{fig:direct_method_comparison}.

Thus,  we can summarize the modified method with the following steps:
\begin{itemize}
\item Start from the frequency surface and compute the gradient to obtain group
velocities.
    
\item Sort the (discrete) gradient by angle and average their magnitudes when
they share the angle.
\end{itemize}

\begin{figure}[h]
    \centering
    \includegraphics[width=4 in]{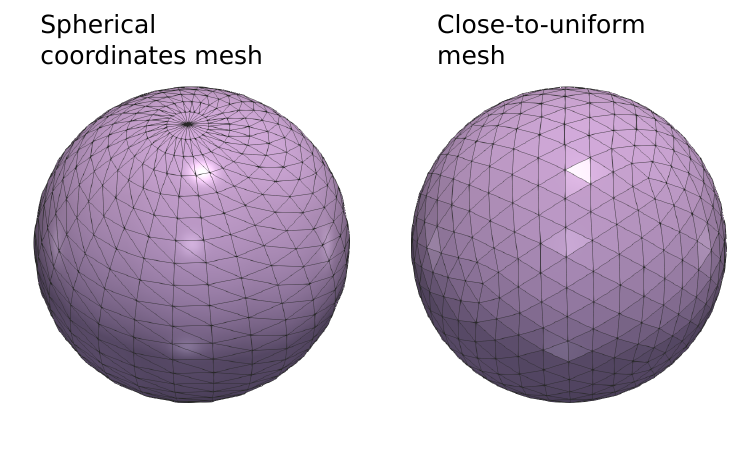}
    \caption{Comparison of a mesh of a sphere using a mesh that follows
    spherical coordinates and a close-to-uniform mesh. Notice the higher
    density of element in the pole of the left mesh.}
    \label{fig:mesh_comparison}
\end{figure}

One natural question that follows is how to extend the method to three
dimensions. But when going from two to three dimensions we do not have a
uniform partition of the sphere. A polyhedron where the vertices of the faces
are uniformly distributed over a spherical surface. In comparison with
two dimensions. Where we used a uniform partition of the circle to sample the
directions for group velocities. In the three-dimensional case, we could
parameterize the sphere using spherical coordinates, for example. This would
represent a problem for our method since it there is a higher density of
polygons in the poles. We expect this because the area differential in spherical
coordinates is given by \(\dd{A} = r\sin\theta\dd{\varphi}\dd{\theta}\)---
here, \(\theta\) is the zenithal angle and \(\varphi\) the azimuthal angle
\cite{book:arfken}. To avoid this problem, we used a triangulated mesh for the
sphere that is close to uniform \cite{meshzoo}. Figure \ref{fig:mesh_comparison}
provides a comparison of two meshes: one using spherical coordinates and the
other with a sampling that is close to uniform.

Another difference that poses a challenge to extend the method from two to three
dimensions is the absence of a \emph{natural} order for the points\footnote{The
preimage of the circle (1D) is  an ordered set, that is, an
interval---that follows the order of the real numbers. We cannot define this
relation directly for the preimage of a sphere (2D).}. To address this issue,
we computed the spherical angles for each point in the sphere and created a k-D
tree \cite{book:effective_computation, scipy}. Then, at the moment of evaluating
the direction of each group velocity vector, we made a nearest-neighbor search
for its angles \((\theta, \varphi)\) to assign the vector to a point on the
sphere as was done in the two-dimensional case.

\pagebreak
\section{Results}

To test the method, we computed the directionality for materials with
different symmetry classes in two and three dimensions. First, we test the
method using homogeneous materials that allow us to (semi) analytically obtain
the dispersion relations and then we try it with results obtained using the
finite element method for a micropolar elastic material. That also shows that
the method does not depend on the equations being analyzed and only needs the
dispersion relations as input.

\subsection{Results for analytic dispersion relations}

We want to solve the determinant in \eqref{eq:christo_freq} for \(\omega\) to
determine the dispersion relations. We rewrite the equation for completeness,
\[\Omega(\omega, \boldsymbol{\kappa}) = \det\left[\Gamma_{ij} - 
\rho \frac{\omega^2}{\Vert\boldsymbol{\kappa}\Vert^2}\delta_{ij}\right] = 0\, .\]
This corresponds to solving a third-degree polynomial equation for each 
wavenumber \(\boldsymbol{\kappa}\) (see the appendix at the end of this chapter for explicit 
forms of these equations). Then, for homogeneous materials, these relations can 
be solved semi-analytically and can be written as
\[\omega \equiv \omega(\boldsymbol{\kappa})\, .\]

On the other hand, when these relationships are obtained from Bloch's theorem
the dispersion relationships also contain information from different Brillouin 
zones leading to relations of the form
\begin{equation}
\omega_{m_1, m_1} \equiv \omega(\boldsymbol{\kappa}_{m_1, m_2})\, ,
\label{eq:disp_general_2d}
\end{equation}
in two dimensions, where the subscripts \(m_1, m_2\) correspond to integer
numbers referring to waves coming from adjacent Brillouin zones. In the case of
three dimensions the relations are of the form
\begin{equation}
\omega_{m_1, m_2, m_3} \equiv \omega(\boldsymbol{\kappa}_{m_1, m_2, m_3})\, .
\label{eq:disp_general_3d}
\end{equation}

In the case of a square/cube unit cell with side \(d\), we have 
the following generalized definition of the wavevector \cite{thesis:langlet}:
\begin{align}
&\boldsymbol{\kappa}_{m_1, m_2} = \left(\kappa_x + \frac{m_1\pi}{d},
\kappa_y + \frac{m_2\pi}{d}\right)\, , \label{eq:wavevector_fbz}\\
&\boldsymbol{\kappa}_{m_1, m_2, m_3} = \left(\kappa_x + \frac{m_1 \pi}{d},
\kappa_y + \frac{m_2\pi}{d},
\kappa_z + \frac{m_3 \pi}{d}\right)\, , \label{eq:wavevector_fbz_3d}
\end{align}
where \(\kappa_x\), \(\kappa_y\), and \(\kappa_z\) are the components of the
wave vector.

\begin{figure}[h]
    \centering
    \includegraphics[width=1.6 in]{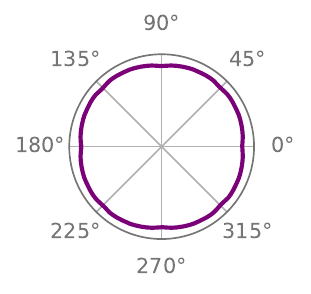}
    \includegraphics[width=1.6 in]{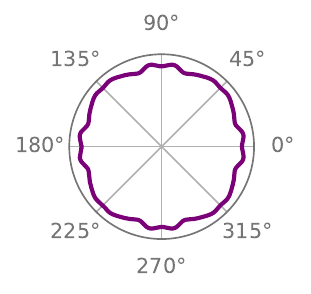}
    \includegraphics[width=1.6 in]{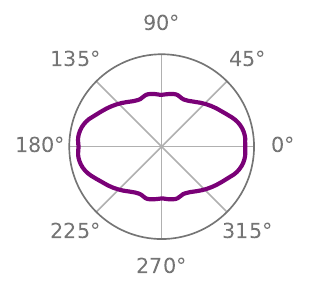}
    \caption{Comparison of directionality for a periodic material with a
    square unit cell for isotropic, cubic, and orthotropic materials.
    We obtained the dispersion curves analytically and restricted them to the
    First Brillouin Zone using \eqref{eq:wavevector_fbz}
    (\textbf{Left}) Directionality for aluminum, an isotropic example.
    (\textbf{Center}) Directionality or GaAs, a cubic example.
    (\textbf{Right}) Directionality for graphite, an orthotropic example.}
    \label{fig:direct2d}
\end{figure}

Figure \ref{fig:direct2d} presents a comparison of directionality for a periodic
material with a square unit cell for aluminum (isotropic), GaAs (cubic) and
Graphite (orthotropic). Here, we obtained the dispersion curves
analytically and restrict them to the First Brillouin Zone using
\eqref{eq:wavevector_fbz}. We can see that the directionality curves present
the same symmetries than the material class in each case (see the appendix at
the end for the material properties). The curve corresponding
to the isotropic material is (almost) symmetric with respect to any rotation in
the plane. In the case of the cubic material, the curve remains the same after
rotations of \(90^\circ\). Finally, for the orthotropic material, we can see
two planes of symmetry corresponding to the \(x\) and \(y\) axes.

\begin{figure}[h]
    \centering
    \includegraphics[width=5.5 in]{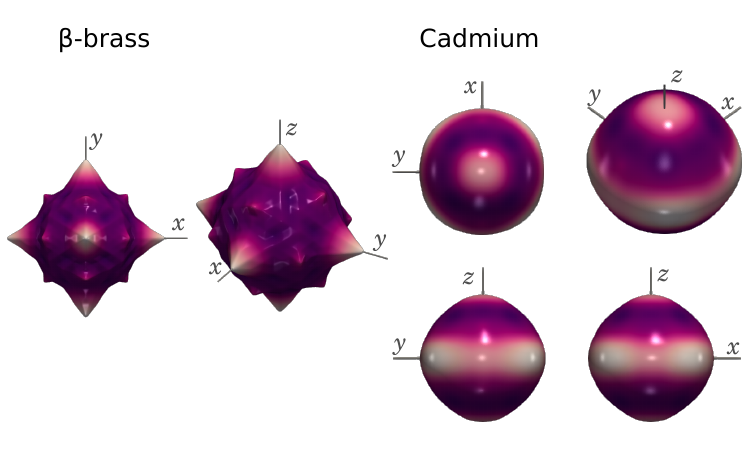}
    \caption{Comparison of the directionality surfaces for (\(\textbf{Left}\))
     \(\beta\)-brass (\(\textbf{Right}\)) and cadmium.}
    \label{fig:direct3d}
\end{figure}

Figure \ref{fig:direct3d} presents a comparison of directionality for a periodic
material made of \(\beta\)-brass (cubic) and cadmium  (transverse isotropic).
For the cubic material figure \ref{fig:direct3d} presents a top view and an
isometric view of the surface, the frontal and lateral views are omitted since
it presents a symmetry with respect to rotations of \(90^\circ\). In the case
of the transverse isotropic material a third-angle projection plus the isometric
view are presented. We can see that the directionality surfaces present the same
symmetries as the material class in each case. In particular, the surface
is (almost) symmetric with respect to the \(z\)-axis (see the appendix at the
end for the material properties used).

\subsection{Results for numerically-obtained dispersion relations}

As a final result, we computed the directionality curves for cellular materials
with circular pores. We changed the diameter of the pore while the size of the
cell is kept fixed. The material used is a micropolar one with the following
properties \cite{guarin-zapata_finite_2020}:
\begin{align*}
&\rho = 2770 \text{ kg/m}^3\, , &\lambda =  5.12\times 10^{10} \text{ Pa}\, ,\\
&\mu = 2.76 \times 10^{10} \text{ Pa}\, , &\alpha = 3.07\times 10^{9} \text{ Pa}\, ,\\
&\gamma + \epsilon = 7.66 \times 10^{10} \text{ N}\, , &J = 306.5 \text{ kg/m}\, .
\end{align*}
where \(rho\) is the mass density, \(\mu\) and \(\lambda\) are the known
Lamé parameters from classical elasticity, while \(J\) is the rotational
inertial density, \(\alpha\), \(\beta\), \(\eta\) and \(\gamma\) are extra
material parameters from the micropolar model and representative of additional
material points interactions see \cite{guarin-zapata_finite_2020, lakes1991,
hassanpour2017} for further discussion on the interpretation of these parameters.

\begin{figure}[h]
    \centering
    \includegraphics[width=5 in]{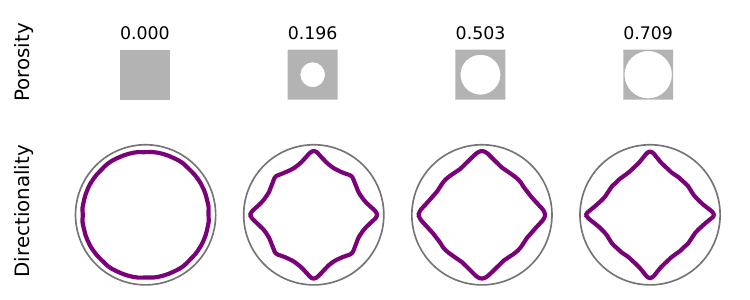}
    \caption{Directionality curves for a cellular material with increasing
    porosity.}
    \label{fig:direct_coss}
\end{figure}

Figure \ref{fig:direct_coss} presents the computed directionality curves for
increasing porosity values, namely: 0.000, 0.196, 0.503 and 0.709. A porosity of
0.0 represents a homogeneous material, used as a reference in this case. As
expected, the directionality of the material increases with porosity, and the
higher values for the averaged group speed occur along the
\(x\) and \(y\) axes where we have continuous paths for the wave to propagate 
\cite{valencia2019}. Notice that the resulting curves are symmetric
with respect to rotations of \(90^\circ\). The same symmetry group can be seen 
in the unit cell.

\begin{figure}[h]
    \centering
    \includegraphics[width=5 in]{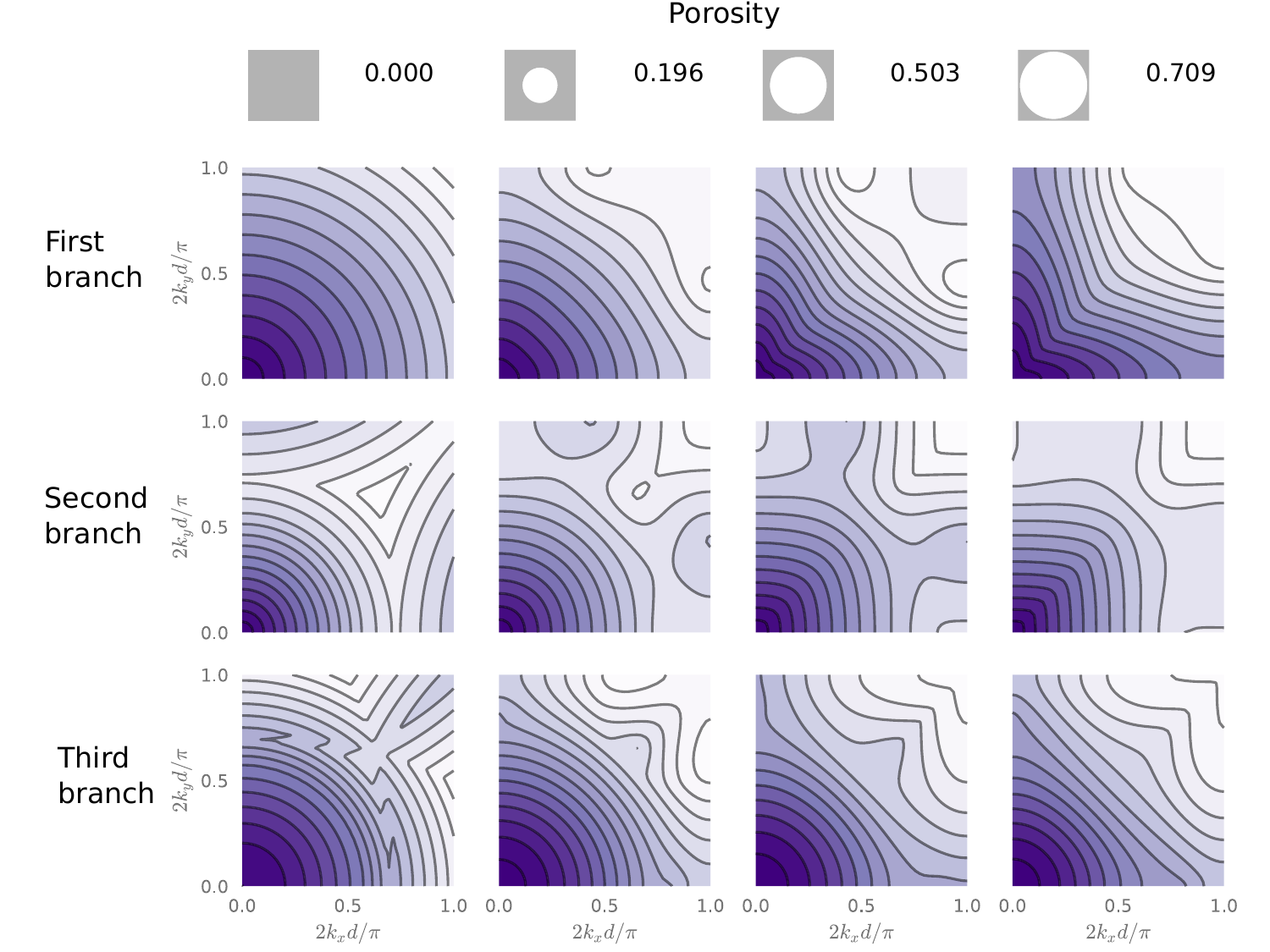}
    \caption{Curves of isofrequency for a cellular material with increasing
    porosity for the first three branches. These are the plots usually used
    to analyze directionality in periodic materials. The
    maximum normalized frequency in each case is
    \(43.38\), \(46.92\), \(57.49\), and \(70.84\)}.
    \label{fig:direct_coss_classic}
\end{figure}

For comparison, we present the isofrequency contours for the
first three branches of the dispersion relations for this material in
Figure \ref{fig:direct_coss_classic} \cite{guarin-zapata_finite_2020}. In this
case we have different normalized frequency ranges --- \(\Omega \in [0, \Omega_{\max}]\)
--- when we change the porosity as the material becomes more dispersive when
heterogeneity appears. Here, the normalized frequency is defined as
\[\Omega = \frac{2d\omega}{c_T}\, ,\]
where \(c_T^2= \mu/\rho\) is the phase speed of the transverse wave in the
low frequency limit for the homogeneous case. The maximum normalized frequency
(\(\Omega_{\max}\)) in each case is \(43.38\), \(46.92\),
\(57.49\), and \(70.84\).

\section{Conclusions}
We presented a method to visualize the directionality of waves in periodic
materials modifying the work of Valencia et al. \cite{valencia2019} and
extending it to the three-dimensional case. This method takes as input
dispersion (hyper-) surfaces obtained using a Bloch analysis and outputs a
curve/surface with the bulk directionality encoded on it. As such, it can be
used for dispersion curves obtained in different physical contexts such as
elastodynamics or electrodynamics. The approach used in this work does not
separate modes \(M_i\) and wave types; instead, it deals with several modes
at once allowing to present the directionality for a broadband frequency range
and not just the low-frequency limit, as is common. Our approach provides a
qualitative tool that is useful to describe the global behavior of waves when
propagating through the analyzed material; it is intended to be used as a
complement to dispersion curves and surfaces.

\bibliographystyle{tfnlm}
\bibliography{directionality_refs}

\appendix

\section{Explicit form for the Christoffel equations}\label{sec:christo}

\subsection{Three dimensions}
For a triclinic material with a stiffness tensor given by
\[[C] =\begin{bmatrix}
C_{11} & C_{12} & C_{13} & C_{14} & C_{15} & C_{16}\\
C_{12} & C_{22} & C_{23} & C_{24} & C_{25} & C_{26}\\
C_{13} & C_{23} & C_{33} & C_{34} & C_{35} & C_{36}\\
C_{14} & C_{24} & C_{34} & C_{44} & C_{45} & C_{46}\\
C_{15} & C_{25} & C_{35} & C_{45} & C_{55} & C_{56}\\
C_{16} & C_{26} & C_{36} & C_{46} & C_{56} & C_{66}
\end{bmatrix}\, ,\]
in Voigt notation, a density \(\rho\) and a wavevector \(\boldsymbol{\kappa}\),
the equation that needs to be solved is \cite{auld1973}
\begin{equation}\label{eq:christo_explicit}
\det\left(\Vert \boldsymbol{\kappa}\Vert^2\begin{bmatrix}
\alpha &\delta &\varepsilon\\
\delta &\beta &\xi\\
\varepsilon &\xi &\gamma
\end{bmatrix}\begin{bmatrix} u_1\\ u_2\\ u_3\end{bmatrix}
- \rho\omega^2\begin{bmatrix}
1 &0 &0\\
0 &1 &0\\
0 &0 &1
\end{bmatrix}\begin{bmatrix} u_1\\ u_2\\ u_3\end{bmatrix}\right) = 0\, ,
\end{equation}
with
\begin{align*}
\alpha =& C_{11} n_1^2 + C_{66} n_2^2 +  C_{55} n_3^2 + 2 C_{56} n_1 n_3 
       + 2C_{15} n_3 n_1 + 2 C_{16} n_1 n_2\, ,\\
\beta =& C_{66} n_1^2 + C_{22} n_2^2 + C_{44} n_3^2 + 2 C_{24} n_2 n_3
       + 2 C_{46} n_3 n_1 + 2 C_{26} n_1 n_2\, ,\\
\gamma =& C_{55} n_1^2 + C_{44} n_2^2 + C_{33} n_3^2 + 2 C_{34} n_2 n_3
       + 2 C_{35} n_3 n_1 + 2 C_{45} n_1 n_2\, ,\\
\delta =& C_{16} n_1^2 + C_{26} n_2^2 + C_{33} n_3^2 + (C_{46} + C_{25})n_2 n_3
       + (C_{14} + C_{56})n_3 n_1\\
       &+ (C_{12}+ C_{66})n_1 n_2\, ,\\
\varepsilon =& C_{15} n_1^2 + C_{46} n_2^2 + C_{35} n_3^2
       + (C_{45} + C_{36}) n_2 n_3 + (C_{13} + C_{55})n_2 n_1\\
       &+ (C_{14} + C_{56})n_1 n_2\, ,\\
\xi =& C_{56} n_1^2 + C_24 n_2^2 + C_{34} n_3^2 + (C_{44} + C_{23})n_2 n_3
       +(C_{36} + C_{45})n_3 n_1\\
       &+ (C_{25} + C_{46})n_1 n_2\, ,
\end{align*}
where \(\hat{\vb{n}} = (n_1, n_2, n_3) = \frac{\boldsymbol{\kappa}}{\Vert\boldsymbol{\kappa}\Vert}\).

In the case of orthotropic materials aligned with the coordinate system these
expressions can be simplified to
\begin{align*}
\alpha =& C_{11} n_1^2 + C_{66} n_2^2 + C_{55} n_3^2\, ,\\
\beta =& C_{66} n_1^2 + C_{22} n_2^2 + C_{44} n_3^2\, ,\\
\gamma =& C_{55} n_1^2 + C_{44} n_2^2 + C_{33} n_3^2\, ,\\
\delta =& (C_{12} + C_{66})n_1 n_2\, ,\\
\varepsilon =& (C_{13} + C_{55})n_3 n_1\, ,\\
\xi =& (C_{44} + C_{23})n_2 n_3\, ,
\end{align*}
this can be further simplified for cubic materials.
\begin{align*}
\alpha =& C_{11} n_1^2 + C_{44}(1 - n_1^2)\, ,\\
\beta =& C_{11} n_2^2 + C_{44}(1 - n_2^2)\, ,\\
\gamma =& C_{11} n_3^2 + C_{44}(1 - n_3^2)\, ,\\
\delta =& (C_{12} + C_{44})n_1 n_2\, ,\\
\varepsilon =& (C_{12} + C_{44})n_3 n_1\, ,\\
\xi =& (C_{12} + C_{44})n_2 n_3\, .
\end{align*}

For materials with transversely isotropic symmetry the equations can be solved
analytically \cite{book:carcione2007}. For a wave propagating in the 
plane 1-3, and taking \(n_2 = 0\), we have
\begin{align*}
\omega_\text{qP}^2 =& \frac{\Vert\boldsymbol{\kappa}\Vert^2 (C_{11}n_1^2 + C_{33}n_3^2
  + C_{55} + \sqrt{M})}{2\rho}\, ,\\
\omega_\text{qS}^2 =& \frac{\Vert\boldsymbol{\kappa}\Vert^2 (C_{11}n_1^2 + C_{33}n_3^2
  + C_{55} - \sqrt{M})}{2\rho}\, ,\\
\omega_\text{S}^2 =& \frac{\Vert\boldsymbol{\kappa}\Vert^2 (C_{66}n_1^2 + C_{55}n_3^2)}{\rho}\, ,\\
M =& [(C_{11} - C_{55})n_1^2 + (C_{55} - C_{33})n_3^2]^2
  + 4[(C_{13} + C_{55})^2 n_1 n_3]^2\, .
\end{align*}

\subsection{Two dimensions}
In the case of a monoclinic material, we could align the symmetry plane to obtain
the following two-dimensional problem
\[\det\left(\Vert\boldsymbol{\kappa}\Vert^2
\begin{bmatrix}\alpha &\delta\\ \delta &\beta\end{bmatrix}
- \rho\omega^2\begin{bmatrix} 1 &0\\ 0 &1\end{bmatrix}\right) = 0\, ,\]
that can be solved analytically as
\[\omega^2 = \frac{\Vert\boldsymbol{\kappa}\Vert^2}{2\rho}[\alpha + \beta \pm
\sqrt{(\alpha - \beta)^2 + 4\delta^2}]\, \]
with
\begin{align*}
\alpha =& C_{11} n_1^2 + C_{66} n_2^2 + 2 C_{16} n_1 n_2\, ,\\
\beta =& C_{66} n_1^2 + C_{22} n_2^2 + 2 C_{26} n_1 n_2\, ,\\
\delta =& C_{16} n_1^2 + C_{26} n_2^2 + (C_{12} + C_{66})n_1 n_2\, ,
\end{align*}
that reduces to
\begin{align*}
\alpha =& C_{11} n_1^2 + C_{66} n_2^2\, ,\\
\beta =& C_{66} n_1^2 + C_{22} n_2^2\, ,\\
\delta =& (C_{12} + C_{66}) n_1 n_2\, ,
\end{align*}
for orthotropic materials and
\begin{align*}
\alpha =& C_{11} n_1^2 + C_{66} n_2^2\, ,\\
\beta =& C_{66} n_1^2 + C_{11} n_2^2\, ,\\
\delta =& (C_{12} + C_{66}) n_1 n_2\, ,
\end{align*}
for cubic materials.

\section{Material properties}\label{sec:materials}
In the following, we present the properties used in the paper.

\subsection{Two dimensions}

\begin{itemize}

\item Aluminum:
\[[C] =\begin{bmatrix}
112.35 & 60.49 & 0\\
60.49 & 112.35 & 0\\
0 & 0 & 25.9\\
\end{bmatrix}\text{ GPa}\,  ,\quad
\rho = 2700\text{ kg/m}^3\, .\]

\item
GaAs:
\[[C] =\begin{bmatrix}
118.8 & 59.4 & 0\\
59.4 & 118.8 & 0\\
0 & 0 & 53.7\\
\end{bmatrix}\text{ GPa}\,  ,\quad
\rho = 5320\text{ kg/m}^3\, .\]

\item
Graphite:
\[[C] =\begin{bmatrix}
235 & 3.69 & 0\\
3.69 & 26 & 0\\
0 & 0 & 28.2\\
\end{bmatrix} \text{ GPa}\, ,\quad
\rho = 1790\text{ kg/m}^3\, .\]

\end{itemize}

\subsection{Three dimensions}

\begin{itemize}

\item Aluminum:
\[[C] =\begin{bmatrix}
112.35 & 60.49 & 60.49 & 0 & 0 & 0\\
60.49 & 112.35 & 60.49 & 0 & 0 & 0\\
60.49 & 60.49 & 112.35 & 0 & 0 & 0\\
0 & 0 & 0 & 25.9 & 0 & 0\\
0 & 0 & 0 & 0 & 25.9 & 0\\
0 & 0 & 0 & 0 & 0 & 25.9
\end{bmatrix}\text{ GPa}\, ,\quad
\rho = 2700\text{ kg/m}^3\, .\]

\item
\(\beta\)-brass:
\[[C] =\begin{bmatrix}
52 & 27.5 & 27.5 & 0 & 0 & 0\\
27.5 & 52 & 27.5 & 0 & 0 & 0\\
27.5 & 27.5 & 52 & 0 & 0 & 0\\
0 & 0 & 0 & 173 & 0 & 0\\
0 & 0 & 0 & 0 & 173 & 0\\
0 & 0 & 0 & 0 & 0 & 173
\end{bmatrix}\text{ GPa}\, ,\quad
\rho = 7600\text{ kg/m}^3\, .\]

\item
Cadmium:
\[[C] =\begin{bmatrix}
115.9 & 41.05 & 41 & 0 & 0 & 0\\
41.05 & 115.9 & 41 & 0 & 0 & 0\\
41 & 41 & 51.2 & 0 & 0 & 0\\
0 & 0 & 0 & 19.95 & 0 & 0\\
0 & 0 & 0 & 0 & 19.95 & 0\\
0 & 0 & 0 & 0 & 0 & 37.43
\end{bmatrix}\text{ GPa}\, ,\quad
\rho = 8650\text{ kg/m}^3\, .\]

\item
Carbon-Epoxy:
\[[C] =\begin{bmatrix}
12.37 & 6.15 & 6.19 & 0 & 0 & 0\\
6.15 & 21.37 & 6.19 & 0 & 0 & 0\\
6.19 & 6.19 & 146.30 & 0 & 0 & 0\\
0 & 0 & 0 & 4.80 & 0 & 0\\
0 & 0 & 0 & 0 & 4.80 & 0\\
0 & 0 & 0 & 0 & 0 & 3.11
\end{bmatrix}\text{ GPa}\, ,\quad
\rho = 1900\text{ kg/m}^3\, .\]

\end{itemize}

\end{document}